\documentclass[fleqn,usenatbib]{mnras}

\usepackage{newtxtext,newtxmath}

\usepackage[T1]{fontenc}
\usepackage{ae,aecompl}

\usepackage{graphicx}	
\usepackage{amsmath}	
\usepackage{amssymb}	

\title[Reverse Shocks in GW GRB counterparts]{Reverse Shocks in the Relativistic Outflows of Gravitational Wave-Detected Neutron Star Binary Mergers}

\author[G. P. Lamb \& S. Kobayashi]{
Gavin P. Lamb$^{1}$\thanks{E-mail: gpl6@le.ac.uk} and
Shiho Kobayashi,$^{2}$
\\
$^{1}$Department of Physics and Astronomy, University of Leicester, University Road, Leicester, LE1 7RH, UK\\
$^{2}$Astrophysics Research Institute, Liverpool John Moores University, IC2, Liverpool Science Park, 146 Brownlow Hill, Liverpool, L3 5RF, UK
}

\date{Accepted XXX. Received YYY; in original form ZZZ}

\pubyear{2019}

\begin{document}
\label{firstpage}
\pagerange{\pageref{firstpage}--\pageref{lastpage}}
\maketitle

\begin{abstract}
The afterglows to gamma-ray bursts (GRBs) are due to synchrotron emission from shocks generated as an ultra-relativistic outflow decelerates.
A forward and a reverse shock will form, however, where emission from the forward shock is well studied as a potential counterpart to gravitational wave-detected neutron star mergers the reverse shock has been neglected.
Here, we show how the reverse shock contributes to the afterglow from an off-axis and structured outflow.
The {off-axis} reverse shock will {appear} as a brightening feature in the rising afterglow at radio frequencies.
{For bursts at $\sim100$\,Mpc, the system should be inclined $\lesssim20^\circ$ for the reverse shock to be observable at $\sim0.1-10$\,days post-merger}. 
For structured outflows, enhancement of the reverse shock emission by a strong magnetic field within the outflow is required for the emission to dominate the afterglow at early times.
Early radio photometry of the afterglow could reveal the presence of a strong magnetic field associated with the {central engine}.
\end{abstract}




\section{Introduction}

The structure of the outflows that drive the shock system responsible for gamma-ray burst (GRB) afterglows is well discussed in the literature \citep[e.g.][]{rossi2002, panaitescu2005, granot2005, salafia2015}.
Due to the highly beamed nature of GRBs, observations of the afterglow are typically limited to cases where the inclination of the system is small and the wider structure of the outflow remains hidden.
However, attempts have been made at interpreting the observational evidence to support various outflow structures in GRBs \citep[e.g.][]{takami2007, pescalli2015, beniamini2019b}.
Gravitational wave (GW) detected mergers involving at least one neutron star will typically be seen off the central rotational axis and will act as a probe for the structure of the jet or outflow that is likely responsible for the cosmological population of short-duration GRBs \citep{lamb2017, lazzati2017, jin2018, kathirgamaraju2018}.
Following the observation via GW of the binary neutron star merger GW170817 \citep{abbott2017}, and year-long observations of the evolving afterglow, constraints on the structure of the afterglow-driving outflow for this event have been made \citep[e.g.][]{gill2018, lamb2018a, lazzati2018, lyman2018, margutti2018, resmi2018, troja2018, vaneerten2018, kathirgamaraju2019, lamb2019}. 

The afterglow estimates for structured outflows have so far ignored the contribution of a reverse shock.
Reverse shocks \citep[e.g.][]{meszaros1997, sari1999, kobayashi2000a, kobayashi2000b, resmi2016} have been identified in the afterglows to long GRBs and should accompany short GRBs, although they have been difficult to detect \citep{lloyd2018}. 
{However, recently the excess radio emission following the short GRB\,160821B has been explained via emission from the reverse shock \citep{lamb2019a, troja2019}.}
The phenomenology of the reverse shock emission can be used as a probe for the magnetisation of the central engine {\citep[e.g.][]{fan2002, zhang2003, zhang2005, giannios2008, gomboch2008, steele2009, mimica2010, granot2012, harrison2013, japelj2014, guidorzi2014, fraija2015, gao2015, kopac2015, zhang2015, huang2016, liu2016, alexander2017, laskar2016, laskar2018, lamb2019a} }and potentially assist in identifying the likely outflow structure.

Constraints on the structure of short-duration GRB outflows have been found following GW170817.
These constraints include:
a narrow jet and high core energy for jet outflows in mergers \citep{beniamini2019}, and a Lorentz-factor for the wider components or cocoon $\geq5$ or $\sim10$ for the cocoon shock breakout scenario \cite[e.g.][]{xie2018, beloborodov2018, fraija2019, matsumoto2019}.
We use these constraints to limit the outflow structure profiles for our reverse shock estimation.

In \S~\ref{sec:model} we discuss the classical reverse shock scenario and in \S~\ref{sec:off} apply the method to the structured outflow models used to produce light-curves for afterglows observed at any inclination.
In \S~\ref{sec:cocoon} we briefly discuss the case of a relativistic cocoon.
In \S~\ref{sec:disc} we discuss our results and in \S~\ref{sec:conc} we give final remarks and conclusions.

\section{Method: the reverse shock}
\label{sec:model}

Using the method for determining the afterglow emission from a structured relativistic outflow in \cite{lamb2017} with the dynamical evolution and expansion description in \cite{lamb2018b} we include synchrotron self-absorption (described below) and add a description for the reverse shock in these systems.
{In this method, the jet/outflow is split into components and the dynamical evolution of each component is treated independently, the emission at equal arrival times from each component is then summed to produce the final light-curve.}
For the reverse shock we follow \cite{kobayashi2000a, kobayashi2000b, harrison2013} and use the dynamical evolution of the blast-wave to scale the reverse shock peak conditions.

The behaviour of the emission from a reverse shock depends, primarily, on the width of the shell through which the shock propagates.
The width of the shell $\Delta_0$ is an unknown free-parameter, although usually assumed to be the product of the speed-of-light $c$ and the GRB duration $T$, giving two cases;
a thick shell, where $\Delta_0 > l/2\Gamma_0^{8/3}$ or thin shell with $\Delta_0 < l/2\Gamma_0^{8/3}$;
here, $l = (3 E_{\rm k}/4\rm{\pi} n m_p c^2)^{1/3}$ is the Sedov length and $\Gamma_0$ is the coasting phase bulk Lorentz factor of the outflow \citep{kobayashi1999}, $E_{\rm k}$ is the isotropic equivalent kinetic energy of the blast-wave, $n$ the ambient number density of protons in the surrounding medium, and $m_p$ the mass of a proton.
{For short GRBs the deceleration timescale is longer than the burst duration $T<l/2\Gamma^{8/3}c$ and so short GRBs are} typically described by the thin shell case.

The synchrotron emission with spectral regime is estimated following \cite{sari1998, wijers1999}.
For a reverse shock in the thin shell case, \cite{kobayashi2000a} demonstrated that the spectral peak flux $F_{{\rm max},r}$, the characteristic frequency $\nu_{m,r}$ and the cooling frequency $\nu_{c,r}$, scale with observed time as $F_{{\rm max},r}\propto t^{3/2}$, $\nu_{m,r}\propto t^6$, and $\nu_{c,r}\propto t^{-2}$ where $t<t_{\rm d}$; and $F_{{\rm max},r}\propto t^{-34/35}$, $\nu_{m,r}\propto t^{-54/35}$, and $\nu_{c,r}\propto t^{4/35}$ where $t>t_{\rm d}$.
Here $t_{\rm d}$ is the observer deceleration time.

For the reverse shock the values of $F_{{\rm max},r}$, $\nu_{m,r}$, $t_{\rm d}$ vary from early analytic estimates via a factor that depends on the dimensionless parameter $\xi_0$, where $\xi_0 = (l/\Delta_0)^{1/2}\Gamma_0^{-4/3}$ \citep{sari1995}.
The correction factors for $F_{{\rm max},r}$ and $\nu_{m,r}$ are defined here as $F_{{\rm max}, r}(t_d)/F_{{\rm max}, f}(t_d) = \Gamma_0~ C_F$, and $\nu_{m,r}(t_d)/\nu_{m,f}(t_d) = \Gamma_0^{-2}~ C_m$, and the observed deceleration time $t_{\rm d} = C_t~l/c~\Gamma_0^{8/3}$.
These correction factors can be approximated as $C_F\sim(1.5+5\xi_0^{-1.3})^{-1}$, $C_m\sim(10^{-2.3}+\xi_0^{-3})$, and $C_t\sim 0.2+\xi_0^{-2}$ respectively \citep{harrison2013}.
As the reverse shock probes the shell material towards the central engine that is driving the outflow, a strong magnetic field associated with the engine will further enhance the reverse shock parameters by a factor $R_B^{1/2}$ for both $F_{{\rm max},r}$ and $\nu_{m,r}$, and by the factor $R_B^{-3/2}$ for $\nu_{c,r}$;
where $R_B\equiv \varepsilon_{B,r}/\varepsilon_{B,f}$ and $\varepsilon_B$ is the magnetic microphysical parameter and the subscript $f$ or $r$ refers to forward or reverse shock respectively \citep{zhang2003, gomboch2008}.
Very high values of $R_B$ have been obtained for some long GRBs \citep{zhang2003, harrison2013, huang2016} {and a value of a $\sim$few for the short GRB\,160821B \citep{lamb2019a}}.

At an inclination $\iota$ that is outside of the jet half-opening angle $\theta_j$, then due to geometric considerations, the observed flux is $F_\nu = F_{\nu, {\rm o}}(\delta/\delta_{\rm o})^k$, where $[\iota-\theta_j]>0$ then $\delta = 1/\Gamma(1-\beta\cos[\iota-\theta_j])$ is the relativistic Doppler factor and $\beta=(1-\Gamma^{-2})^{1/2}$, and the subscript `o' indicates the on-axis value $\delta_{\rm o}=1/\Gamma(1-\beta)$.
The value of $k$ depends on the separation from the jet edge with $k\sim 2$ for $\iota\lesssim2\theta_j$ and $k\sim3$ for $\iota\gtrsim2\theta_j$ \citep{ioka2018}.
For an outflow with angular structure\footnote{Angular structure refers to an outflow with energy and/or Lorentz factor that vary with angular separation from the central axis; $[E_{\rm k}(\theta), \Gamma_0(\theta)] \propto f(\theta)$} we sum the evaluated flux from each angular segment across the outflow.

At low frequencies synchrotron self-absorption (SSA) becomes
important.
SSA limits the flux for the reverse shock more efficiently than for the forward shock due to the lower effective temperature of the electrons in the reverse-shock region.
The limiting flux, at a given frequency $\nu$ and observer time $t$, in the reverse shock can be estimated by considering the intensity of a black-body with the reverse shock temperature \citep[e.g.][]{kobayashi2000b, nakar2004}.
\begin{equation}
    F_{\rm BB} \sim 2m_p~(1+z)^3~\delta~\varepsilon_e~\nu^2\frac{p-2}{p-1}~\frac{e}{\rho}\left(\frac{R}{D_L}\right)^2~\Omega~\cos{\theta}~{\rm max}\left[\frac{\nu}{\nu_m}, ~1\right]^{1/2},
\end{equation}
where, $z$ is the redshift, $\varepsilon_e$ is the {fraction of the shock energy that is partitioned to electrons}, $\nu$ is the observed frequency, $e$ is the internal energy density, $\rho$ the mass energy density, $R$ is radius of the blast-wave, $D_L$ the luminosity distance, and $\Omega$ and $\theta$ are the solid angle and opening angle of the emission region\footnote{We split the jet into different emission regions defined by a solid angle $\Omega$ and an opening angle $\theta$ where $\Sigma_i\Omega_i = \Omega_j\sim \pi\theta_j^2$}.  
Here the ratio $e/\rho$ is $\sim8\times10^{-2}$ \citep{harrison2013} until the shock crossing time where it evolves as $t^{-2/7}$.
Alternatively, see \cite{resmi2016} where they consider the opacity of the source to estimate the SSA limit.

\subsection{Reverse shocks viewed off-axis}
\label{sec:off}

\begin{figure*}
	\includegraphics[width=\textwidth]{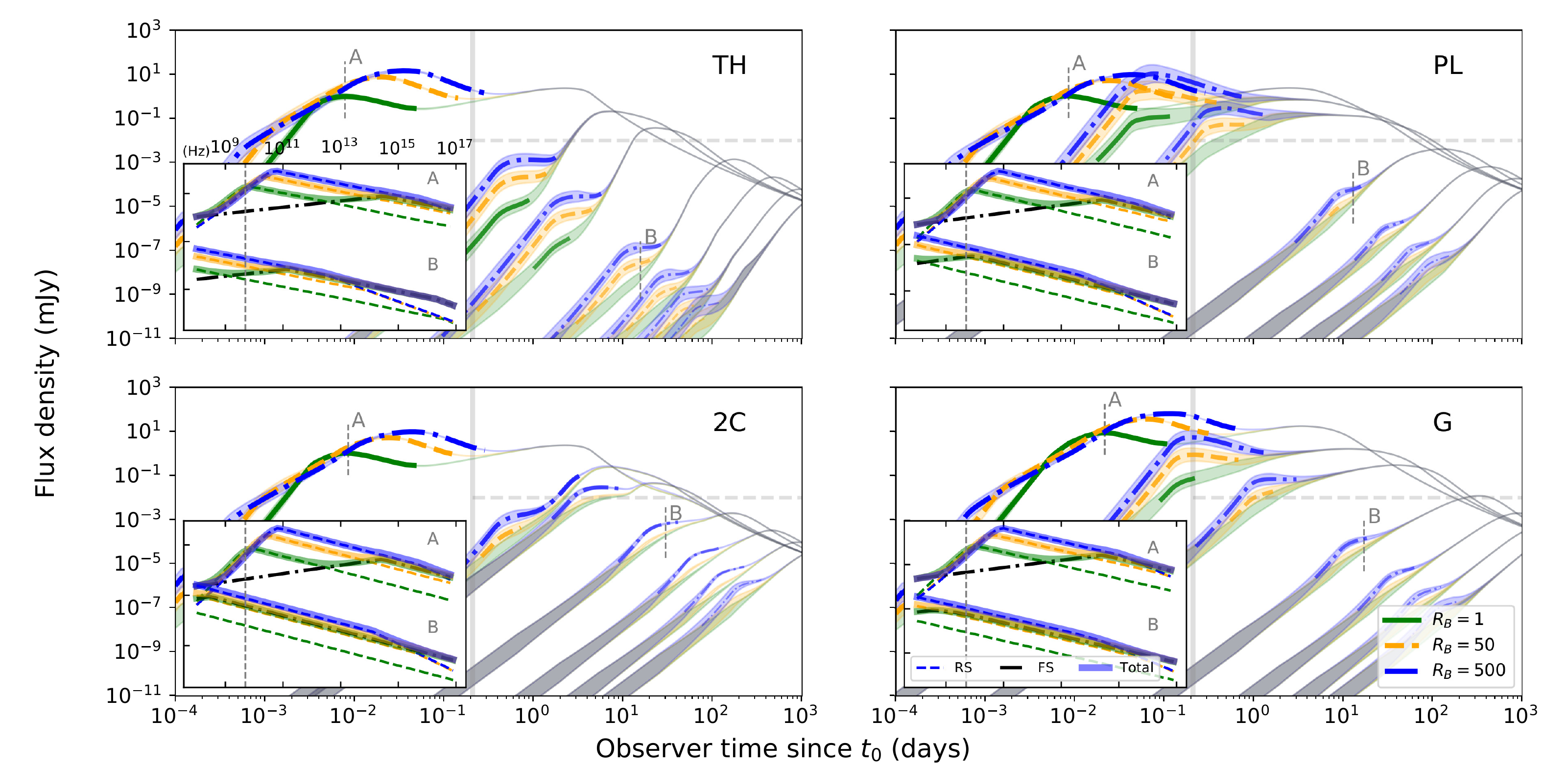}
    \caption{Jet structure afterglows at a distance 100 Mpc and observed in radio at 5\,GHz.
    Afterglows are viewed at $[0,~12,~18,~36,~54,~72,~90]^\circ$ where $0^\circ$ is indicated by the light-curve denoted `A', and all all subsequent light-curves are for increasing inclination.
    Three magnetization parameters are shown, $R_B=[1, 50, 500]$, solid green, dashed orange, and dash-dotted blue line respectively. 
    The four structure models are as described in the text:
    top left -- `top hat' (TH); bottom left -- two-component (2C); top right -- power-law (PL); bottom right -- Gaussian (G).
    The $x$-axis shows the time since $t_0$ -- either, a GRB trigger for an on-axis case, or a GW trigger when off-axis. 
    {The light-curve is shown with the colour that corresponds to the $R_B$ parameter where the afterglow is reverse shock dominated, and in grey where the forward shock dominates.
    The uncertainty in the flux due to scintillation is shown as a shaded region while the source size is small.
    The vertical grey line shows the 5 hour post merger/GRB and is representative of the earliest time the VLA can be observing.
    The horizontal dashed grey line indicates 10$\mu$Jy, the $\sim$sensitivity limit for a 1 hour exposure \citep{perley2011} and the limit at which GRB radio afterglows have been detected \citep{macpherson2017}.
    The sub-plot in each panel shows the spectral energy distribution (SED) corresponding to the on-axis light-curve at the time indicated by a vertical grey dashed line (in the main panel) and the letter `A' and the SED at $36^\circ$ and marked with a `B'.
    The black dashed-dotted line in the SED is the forward shock contribution while the dashed line represents the RS with a given $R_B$.
    The vertical grey dashed line in the SED indicates 5 GHz.
    }}
    \label{fig:TH3GHz}
\end{figure*}

\begin{figure*}
	\includegraphics[width=\textwidth]{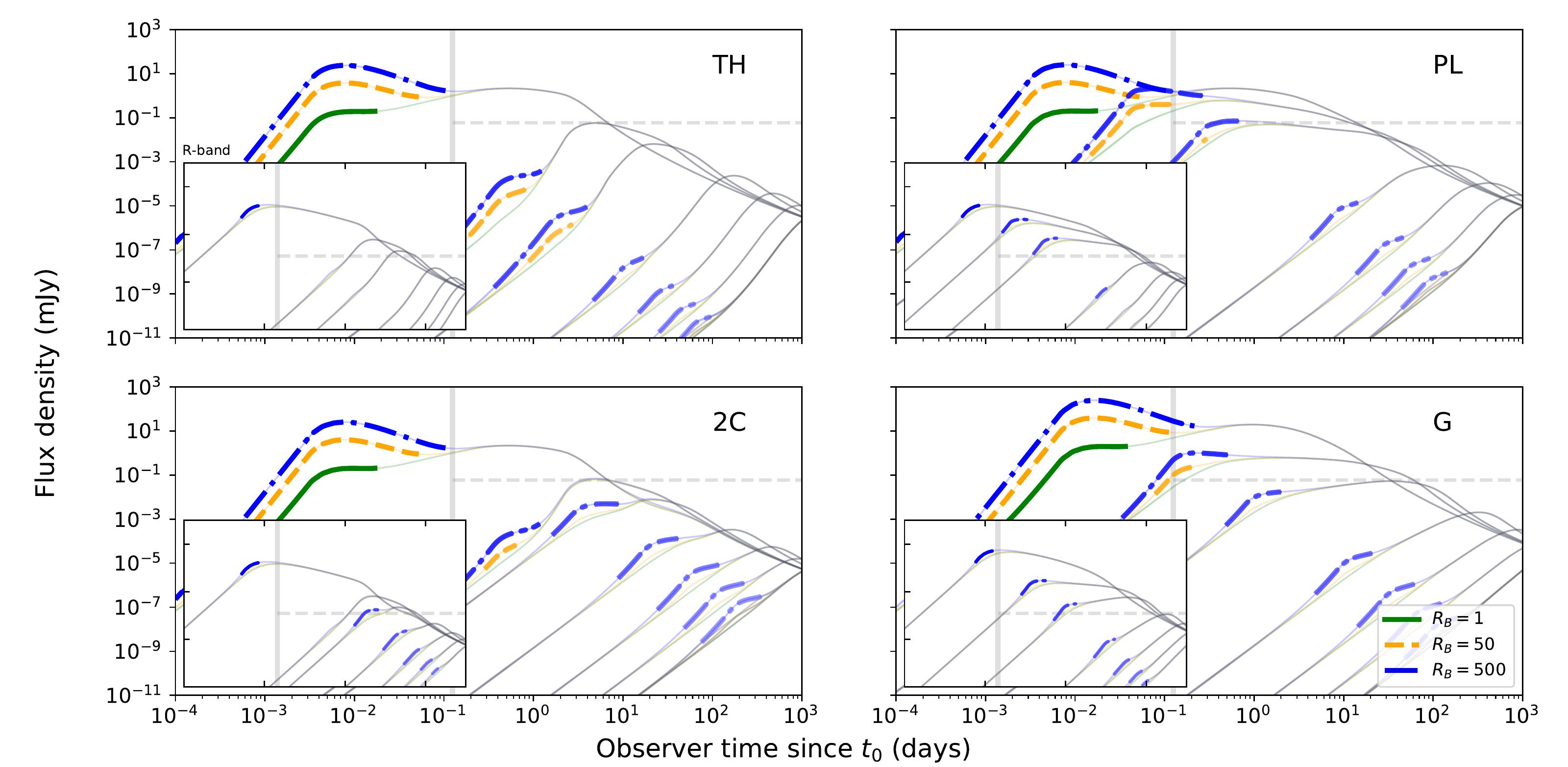}
    \caption{{Light-curves as in Figure\ref{fig:TH3GHz} but at 97.5 GHz and R-band (inset). Vertical line in main panels indicates 3 hours post merger and the horizontal dashed line indicates 60\,$\mu$Jy, the response and $\sim$sensitivity limit of ALMA at 97.5\,GHz \citep[e.g.][]{macpherson2017}. The vertical line in the inset indicates 30 minutes post-merger and the horizontal dashed line shows $m_{{\rm AB}}=21$}.}
    \label{fig:THalma}
\end{figure*}

To determine if the signature of a reverse shock is apparent in the afterglow from a GW-detected merger jet we estimate the flux from a variety of outflow structures with a range of inclinations.
Following \cite{lamb2017} we consider four jet structures generally described as a top-hat, two-component, power-law, and a Gaussian.
The top-hat model is a jet with a uniform kinetic energy and velocity distribution and sharp edges at the value $\theta_c$ used as the core angular width for the jets with a more complex angular structure.
The two-component model follows \cite{lamb2019} with a top-hat jet surrounded by a second component with 10\% of the core isotropic equivalent energy and a Lorentz factor of 5.
For the power-law model we follow \cite{lamb2017} where outside of a top hat core the energy and Lorentz factor scale with angle as $\propto(\theta/\theta_c)^{-2}$, and a condition ensuring $\Gamma\geq1$.
The Gaussian model follows the description in \cite{lamb2018a, resmi2018, lamb2019}; $E(\theta)=E_{\rm c}~e^{-\theta^2/\theta_c^2}$ and $\Gamma_0(\theta)= (\Gamma_{0,{\rm c}}-1)~e^{-\theta^2/2\theta_c^2}+1$, where the subscript `c' indicates the central or core values.
For all the structured jets we limit the structure by imposing an edge at $\theta_j=15^\circ$.

We fix various fiducial parameters for the jets with the core, or central values, as $E=10^{51}$ erg (or $10^{52}$ erg for the Gaussian model), $\Gamma_0=100$, and $\theta_c=6^\circ$.
The other model parameters are;
the electron distribution index $p=2.2$, the microphysical parameters $\varepsilon_{B,f}=\varepsilon_e^2=10^{-2}$, and the ambient density $n=10^{-3}$ cm$^{-3}$.

Fig.~\ref{fig:TH3GHz} shows the afterglow light-curves, observed at 5 GHz and inclinations $\iota=[0,~2\theta_c,~3\theta_c,~6\theta_c,~9\theta_c,~12\theta_c,~15\theta_c]$, for the four jet structure models considered\footnote{We do not consider the counter-jet here and therefore the flux density at $\sim90^\circ$, or our $15\theta_c$, will be brighter by a factor $2$ where the counter-jet is identical to the forward jet}.
The reverse shock, in each case, peaks for an on-axis observer at $t\sim0.001 - 0.1$ days.
The second peak at $\sim1 - 10$ days is the forward shock afterglow.
{The light-curve in Fig.~\ref{fig:TH3GHz} is coloured according to the value of the magnetization parameter $R_B$ while the afterglow is dominated by the reverse shock.}
For an off-axis observer the reverse shock is expected to contribute before the afterglow peak time although, in some cases for a structured jet, the reverse shock can result in a two peaked afterglow.

{The effects of scintillation on the observed flux have been considered.
Scintillation is most apparent at low-frequencies, typically $<10$\,GHz, and at early times when the source is compact.
As the jet expands, the size of the source increases and the effects of scintillation are reduced.
We estimate the size of the outflow at each time-step by considering the angle subtended by the emitting surface, the inclination to the line-of-sight, and the radius of the blast-wave.
The size on the sky is then estimated considering the distance to the source.
Following \cite{walker1998,walker2001, granot2014} we can use the angular size of the first Fresnel zone $\theta_{\rm FO}=6.32\times10^4 SM^{0.6}\nu_0^{-2.2}~\mu$as, where $SM$ is the scattering measure and $\nu_0$ the transitional frequency, to estimate the modulation index $m$ for the relevant scintillation regime.
From the NE2001\footnote{\url{https://www.nrl.navy.mil/rsd/RORF/ne2001/##los}} model \citep{cordes2002}, the typical values are $SM\sim10^{-3.5}$\,kpc/m$^{20/3}$ and $\nu_0\sim10$\,GHz.}

The afterglows in Fig.~\ref{fig:TH3GHz} show the reverse shock is self-absorbed before the peak time in all {on-axis $\iota\sim0^\circ$} cases i.e. $F_{\rm BB}<F_{\nu}$.
{This is highlighted in the insets, where we show the spectral energy distribution (SED) at times marked by the letter `A' and `B' respectively are shown.
The contribution of the reverse shock is shown as dashed and coloured lines (according to the value of $R_B$) and the contribution from the forward shock is shown as a black dash-dotted line, the solid coloured lines indicated the sum of the two components.
For the on-axis case `A', the peak of the SED indicates the self-absorption frequency where $F_{\rm BB}(\nu)=F_\nu$, this is consistent with values for the self-absorption frequency in \cite{nakar2004}.}
For an off-axis observer, the SSA emission has a limited contribution {and the typical SED is a single power-law from radio to X-ray frequencies, however,}  SSA effects can be seen for the power-law structured case where the magnetization is high and the system mildly inclined $\iota=12^\circ$ {near the reverse shock peak}.

{The light-curve at 97.5\,GHz and $R$-band (inset) are shown in Fig.~\ref{fig:THalma}, where the reverse shock dominates the afterglow, the light-curve is shown with a coloured line.
At these higher frequencies scintillation has no effect and self-absorption is not apparent consistent with the SED in Fig.~\ref{fig:TH3GHz}.}
For the top-hat jet (TH), top-left in Figs.~\ref{fig:TH3GHz} {and \ref{fig:THalma}}, the off-axis reverse shock emission results in a brightening feature in the rising afterglow.
For a magnetized ejecta where $R_B>1$ this feature is present at {$\sim5$\,GHz for all inclinations.
Where $R_B\gtrsim500$, the reverse shock feature is present at $\sim97.5$\,GHz, and for structured jets at optical frequencies.}

The two-component (2C), shown bottom-left in Fig.~\ref{fig:TH3GHz} {and \ref{fig:THalma}}, shows similar features to the TH case {at $\iota\lesssim12^\circ$}.
However, at higher inclinations, the forward shock emission from the low-$\Gamma$ wide component competes with the off-axis emission from a reverse shock in the jet core.
A reverse shock in the wider component is faint and only appears brighter than the forward shock afterglow where $R_B>1$, {and can be seen at inclinations $\iota>12^\circ$}.
At higher inclinations, the reverse shock emission from the low-$\Gamma$ wide/second component results in an afterglow that rises to a plateau, for $R_B=500$, before the forward shock emission from the energetic core dominates and results in the late-time peak.

The right panels in Fig.~\ref{fig:TH3GHz} {and \ref{fig:THalma}} show the afterglow light-curves for a power-law (PL) and a Gaussian structured (G) jet (top and bottom respectively).
Phenomenologically, these two cases appear similar;
the smooth change in the energy and Lorentz factor profile means that, overall, afterglow emission is brighter for longer for an off-axis observer than for a regular top-hat jet -- this is consistent with the findings of \cite{lamb2018c} where orphan afterglows from structured jets have a higher rate of two or more detections with typical survey telescope cadences.
The reverse shock for the highly magnetized cases are observable above the forward shock, even at high inclinations, {where the { emission is dominated by the} reverse shock in the lower energy wider components of the outflow structure}.
For both the power-law and Gaussian structured jets, observed at mild inclinations $\iota\sim(3-4)\theta_c$ (up to $\sim6\theta_c$ for the Gaussian case), the reverse shock peak coincides with the beginning of the characteristic flat or shallow rise to peak i.e. the $t^{4/5}$ incline observed in the pre-peak afterglow to GRB\,170817A \citep[e.g.][]{lyman2018, lamb2019}.

\subsection{Reverse shocks in cocoons}
\label{sec:cocoon}

\begin{figure}
	\includegraphics[width=\columnwidth]{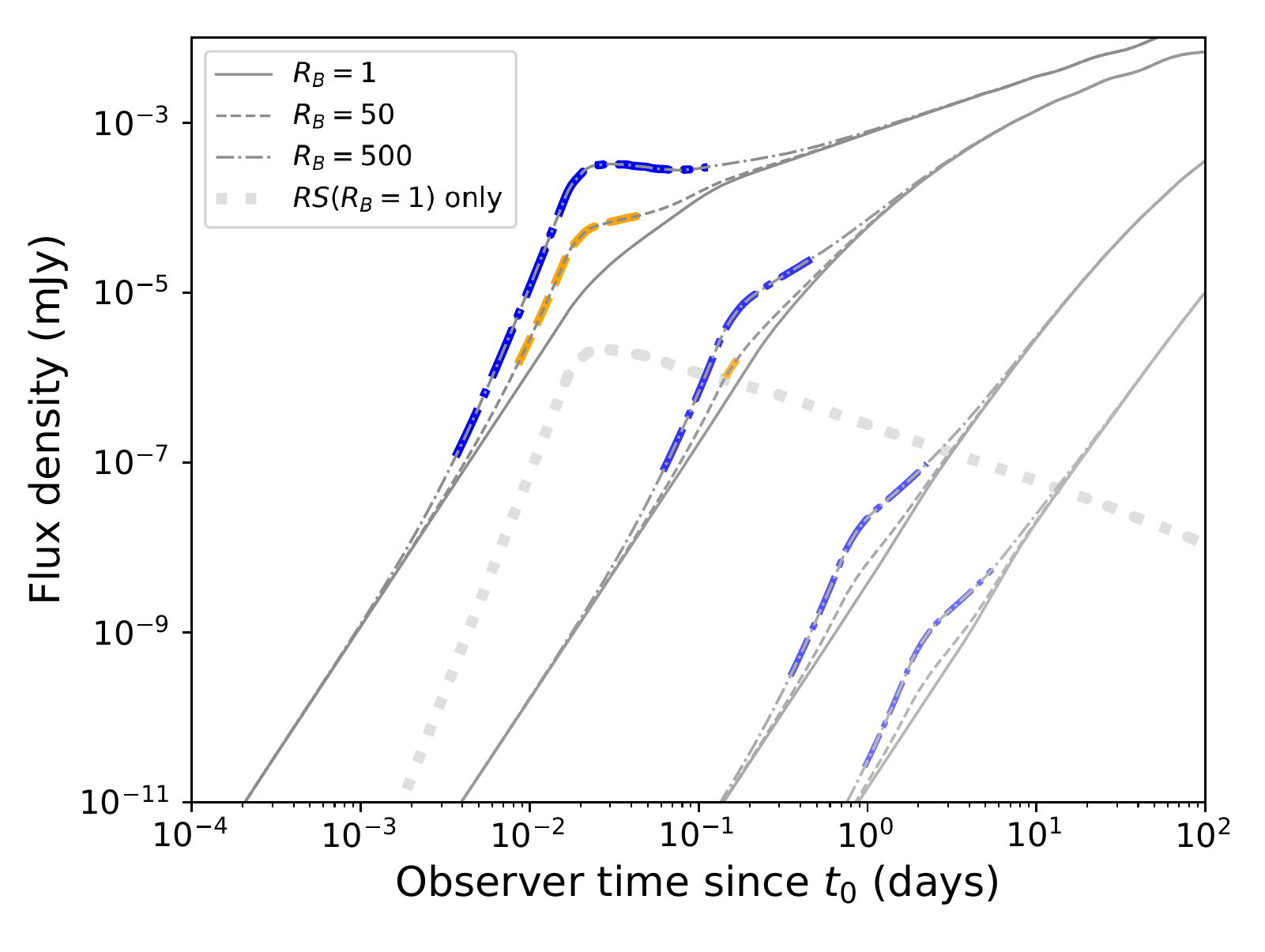}
    \caption{{A choked-jet cocoon afterglow at 5\,GHz viewed at $\iota=[0^\circ,45^\circ,70^\circ,~{\rm and}~90^\circ]$. The contribution to the afterglow from a reverse shock with $R_B=[1,50,500]$ is shown in each case. The component of the flux from the reverse shock is shown for the on-axis and $R_B=1$ case as a thick dotted grey line. From this it is clear to see that the reverse shock never contributes significantly to the emission.}}
    \label{fig:cc}
\end{figure}

A jet that stalls as it drills through the envelope of material ejected during the merger process will inflate a cocoon of energised matter \citep{murguia2014, murguia2017, gottlieb2018a}.
As the cocoon material propagates into the surrounding medium it will sweep up material in the same fashion as a GRB jet, generating a shock system that will produce a broadband afterglow.
For such a cocoon of material, with a relativistic velocity distribution, the slower components will catch up and refresh the forward shock creating a distinctive, slow rising afterglow \citep[e.g.][]{mooley2018a}.
Although the afterglow following GRB\,170817A was not due to such a choked-jet system, such transients may exist and the electromagnetic counterparts to future GW detected mergers may reveal the afterglow to a choked-jet cocoon.
The existence of such a choked-jet population is supported by the duration analysis of short GRBs \citep{moharana2017}.

The afterglow from a wide-angled choked-jet system will be semi-isotropic, depending on the initial opening angle of the outflow $\theta\sim 30-40^\circ$ \citep{nakar2018a}, and potentially {broadband detectable} on long timescales ($\sim100$s days) for nearby {($\sim 50$\,Mpc)} events where the cocoon has a radial velocity distribution \citep{fraija2019}.
A reverse shock will travel back into the cocoon and the forward shock will be continuously refreshed and energised \citep{sari2000}.
The reverse shock probes the slower material catching and energising the forward shock system.
While slower material continues to refresh the forward shock, the reverse shock will persist \citep{rees1998, sari2000}.

The maximum synchrotron flux from a reverse shock for a cocoon with a uniform energy distribution and a fastest component with $\Gamma_0=10$ will be $F_{{\rm max}, r} \sim R_B^{1/2}~\Gamma_0~C_F~F_{{\rm max}, f}$;
in the thin shell regime\footnote{{The assumption of a thin shell is due to the nature of the reverse shock in an outflow with a radial velocity distribution; as the slower material catches the decelerating forward shock a reverse shock forms, as the shock system is continuously energised by the slower arriving material the reverse shock can be instantaneously approximated by an infinitesimally thin shell of shocked material.}} where $\xi_0>>1$ then $C_F\rightarrow 0.667$.
Similarly, the coefficient for the characteristic frequency $\nu_{m, r}$ is $C_m\rightarrow 5\times10^{-3}$.
As the slower components refresh the system, the relevant Lorentz factor $\Gamma_0$ for the reverse shock will be reduced -- $F_{{\rm max}, r}$ and $\nu_{m,r}$ depend on the Lorentz factor as $\Gamma_0$ and $\Gamma_0^{2}$ respectively.

For the fastest component in a system with our fiducial parameters {(e.g. $p=2.2$,  $\varepsilon_B=0.01$, $\varepsilon_e=0.1$ and $n=0.001$\,cm$^{-3}$)}, the forward shock will have a characteristic frequency, at the deceleration time, $\nu_m\sim 3.2\times10^{10}$ Hz and a peak synchrotron frequency for the reverse shock at $\sim 1.6\times10^6$ Hz;
assuming slow cooling, the flux at $\sim5$\,GHz would be a factor $\sim 0.1$ of the peak forward shock flux at $t_{\rm d}$.
For a velocity distribution within the cocoon that ranges from $\Gamma = 10 - 1.4$, then as the forward shock is energised, where $E(>\Gamma\beta)\propto(\Gamma\beta)^{-\kappa}$ here $4.5\leq\kappa\leq6.2$ \citep{nakar2018b}, and $F_{{\rm max,f}}\propto E$ the forward shock emission will always dominate over that from the reverse shock which propagates into a shell with the lower initial energy.

In the case where the cocoon is magnetized, the reverse shock for our parameters will initially dominate over the forward shock, {where $\nu<\nu_{m,f}$ for the forward shock and $\nu>\nu_{m,r}$ for the reverse shock, then the minimum magnetization parameter for $F_{\nu,r}/F_{\nu,f}>1$ is $R_B>[\Gamma_0^{2-p}C_FC_m^{(p-1)/2}(\nu/\nu_{m,f})^{(1-3p)/6}]^{-4/(p+1)}$ giving} $R_B\gtrsim18$ {for our typical parameters}.
Where the initial Lorentz factor of the outflow is $<10$ the required $R_B$ increases\footnote{Where $\nu_{m,f}\propto\Gamma_0^4$ at $t=t_{\rm d}$, and $R_B$ is then proportional to a negative power of $\Gamma_0$.} e.g. for $\Gamma_0=7$, then $R_B\gtrsim86$.
As the forward shock is energised, the emission from the reverse shock will be buried beneath that from the forward shock.
In such a case, the signature of a reverse shock, will appear as a radio flare at $\sim t_{\rm d}$ for the outflow.
{Fig.~\ref{fig:cc} shows the afterglow from a choked-jet cocoon at 50\,Mpc with a $\kappa=6$ and observation angle $\iota=[0^\circ,45^\circ,70^\circ,90^\circ]$ and the reverse shock with $R_B=[1,50,500]$.
For the case where $R_B=1$, the reverse shock never dominates emission over the forward shock; the dotted grey line indicates the reverse shock contribution at $\iota=0$ and $R_B=1$.
The slow decline in the reverse shock emission post-peak traces the Lorentz factor of the radial velocity distribution $\Gamma=10 \rightarrow 1.4$.}

\section{Discussion}
\label{sec:disc}

\begin{figure*}
	\includegraphics[width=\textwidth]{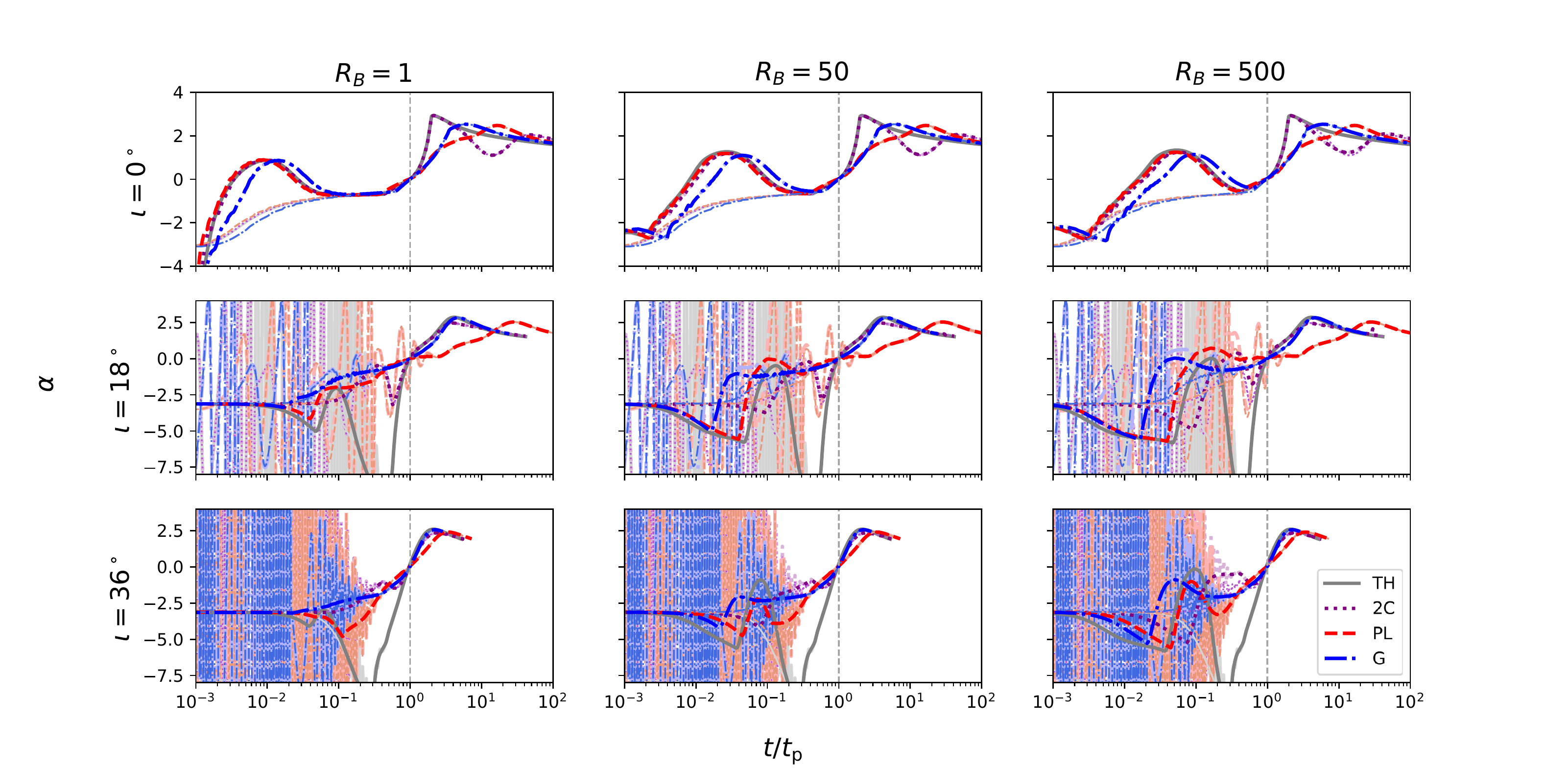}
    \caption{The rise index $\alpha$ defined as $F\propto t^{-\alpha}$ for the 5\,GHz afterglow light-curves at an inclination $\iota=[0^\circ,~18^\circ,~{\rm and}~36^\circ]$ or $[0,~3,~{\rm and}~6]\times\theta_c$.
    { The thin lines show $\alpha$ for the forward shock only case and the faint lines show the effects of refractive and diffractive scintillation, which are particularly prominent at early times, see text in \S \ref{sec:disc}.}
    The $x$-axis shows time normalised to $t_{\rm p}$, the observed light-curve peak due to emission from the forward shock, the vertical line at $t/t_{\rm p}=1$.
    Each column shows a single magnetic parameter, $R_B=[1,50,500]$ respectively.
    The jet structures are indicated by the line colour and style:
    TH -- top-hat with a solid grey line;
    2C -- two-component with a dotted purple line;
    PL -- power-law with a dashed red line;
    and G -- Gaussian with a blue dashed-dotted line.}
    \label{fig:alpha3GHz}
\end{figure*}

\begin{figure}
	\includegraphics[width=\columnwidth]{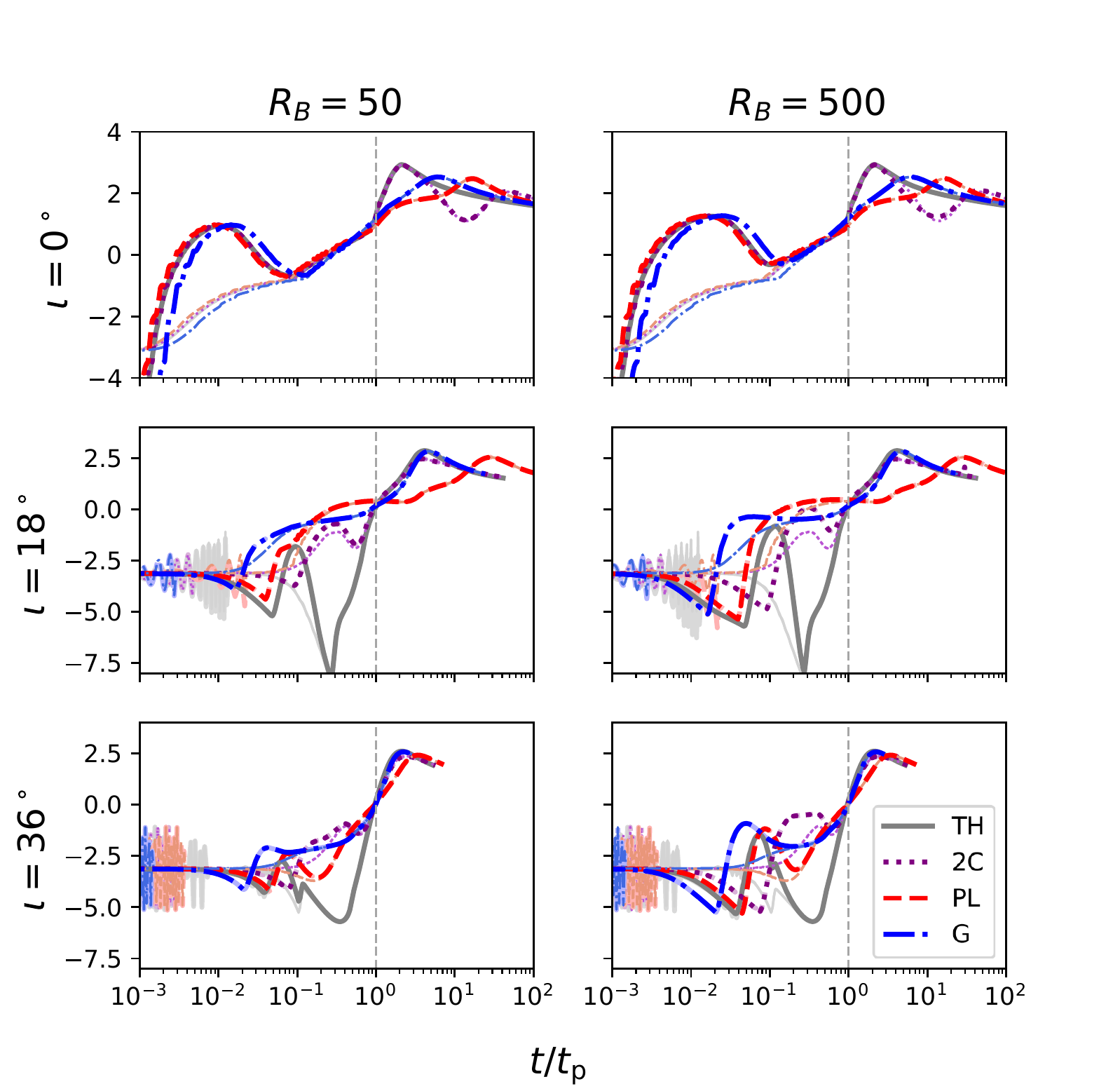}
    \caption{Same as Fig. \ref{fig:alpha3GHz} but showing the rise index $\alpha$ (y-axis) with forward shock normalised time $t/t_{\rm p}$ at 97.5\,GHz, (ALMA). The effects of scintillation, shown as faint lines, are much reduced due to the higher frequency. The rise index for the forward shock only case is shown as a thin line.}
    \label{fig:alphaALMA}
\end{figure}

\begin{figure}
	\includegraphics[width=\columnwidth]{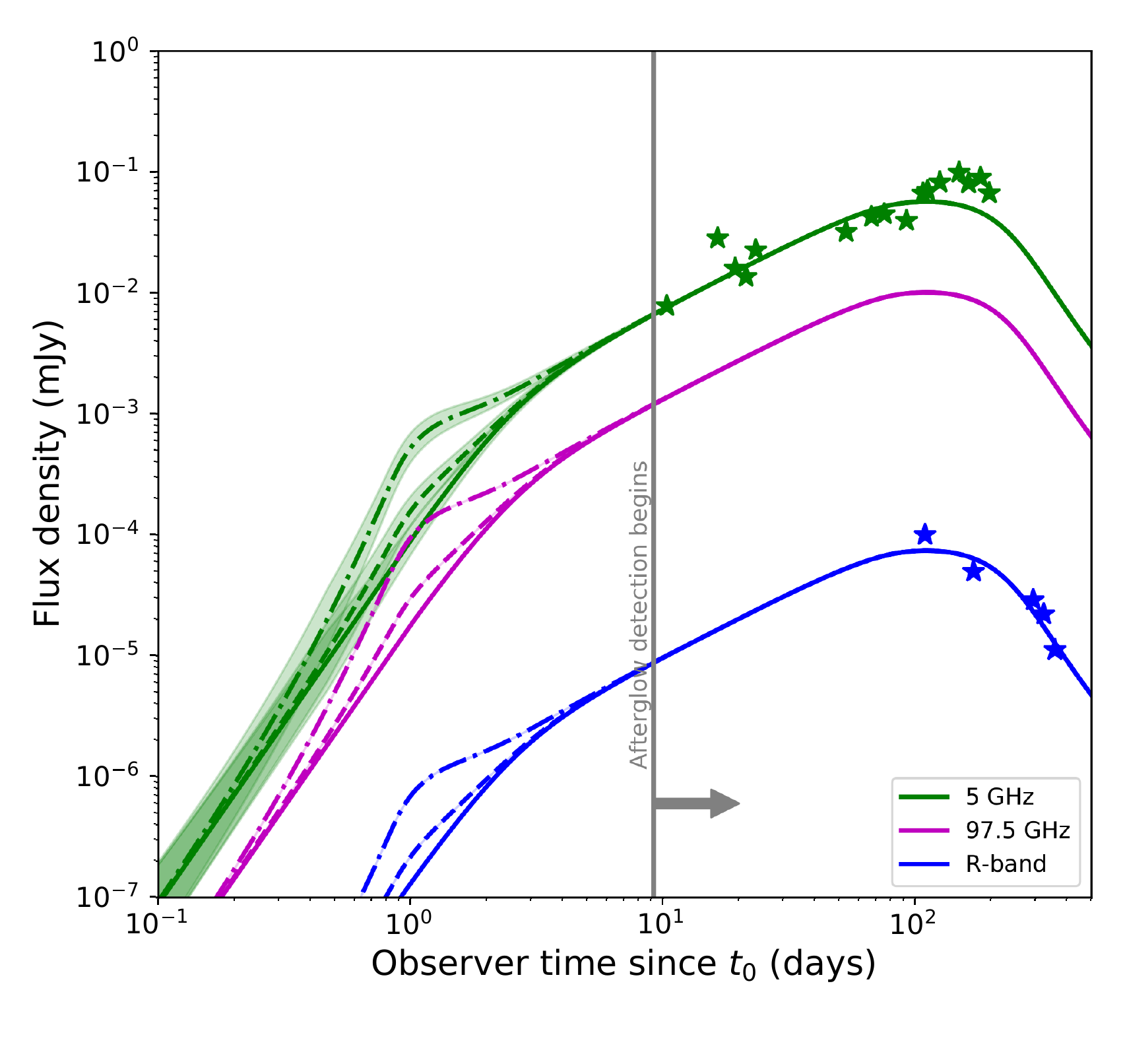}
    \caption{{Data of the afterglow to GW170817 at $\sim5$\,GHz and $R$-band \citep{lyman2018,lamb2019,dobie2018,hallinan2017, margutti2018, mooley2018a} and a light-curve at 5\,GHz, 97.5\,GHz and $R$-band with parameters typical of GRB\,170817A fits \citep[e.g.][]{lamb2019}. The reverse shock for such a structured jet is apparent at $\sim1$\,day post merger. The solid, dashed, and dashed-dotted lines indicate an $R_B=[1,50,500]$ respectively. The reverse shock is dominant at $\sim1$\,day only for cases where $R_B>1$.}}
    \label{fig:GW170817}
\end{figure}

By considering a reverse shock for various jet or outflow structures, we have shown that the pre-peak afterglow for an off-axis observer will contain a distinctive feature with a reverse shock origin.
A larger residual magnetic field from the central engine will enhance the reverse shock emission, and for small inclinations, may result in the brightest afterglow peak when observed at low-frequencies.

For short-duration GRBs, the low characteristic frequency and the early peak time for the reverse-shock emission means that fast response and deep radio photometry of GW triggered neutron star mergers is critical in identifying the reverse shock contribution {\citep[we note that the same criteria apply to the cosmological sample of short GRBs that are typically at $z=0.5$][]{berger2014}}.
At a distance of 100\,Mpc, where a system is inclined $\iota<20^\circ$ the reverse shock feature will appear on a timescale $t\sim0.001 - 10$\,days post merger.
{For such systems we can expect a high energy trigger, either an X-ray flash or a GRB;
where such emission is observed by {\it Swift}/Burst Alert Telescope (BAT) then the source will be easily localised \citep[see text and Figure 1 in][]{mandhai2018} and rapid follow-up can commence.
}

{The vertical lines in Fig.~\ref{fig:TH3GHz} indicate 5 hours post-merger and the typical response time of the Karl-Jansky Very Large Array (VLA).
The horizontal dashed line indicates the typical sensitivity limit of $10$\,$\mu$Jy \citep{perley2011, macpherson2017}.
Similar vertical lines are shown in Fig.~\ref{fig:THalma} where at 97.5\,GHz they represent the response and sensitivity of ALMA, $\sim3$\,hours and $\sim60\mu$Jy \citep[e.g.][]{macpherson2017}, and for the inset the $R$-band response $\sim0.5$\,hours and a magnitude $\sim21$ -- at optical frequencies the response of various facilities can be within seconds of receiving a trigger and the limiting magnitude can vary from telescope-to-telescope. 
From these limits it is clear that radio follow-up of GW detected mergers should focus on nearby $<100$\,Mpc and mildy inclined $\iota\lesssim20^\circ$ (where information is available) and in all cases where the merger is accompanied by a high energy electromagnetic trigger.
We additionally note that the timescales shown here depend on the ambient density, energy, and Lorentz factor of the outflow - crucially we note that where $\Gamma<100$ the reverse shock peak will appear later, and where $E>10^{51}$\,erg and $n>10^{-3}$\,cm$^{-3}$ then the peak afterglow flux will be brighter and the timescale longer for a higher energy outflow and shorter for a higher density environment.
For GRB\,160821B, early radio observations at 0.15\,days show the reverse shock and require an outflow with $\Gamma\sim60$ \citep{lamb2019a}}.

Our example in Fig.~\ref{fig:TH3GHz}, shows that for an observer at the typical GW detected inclination of $\sim38^\circ$ \citep{lamb2017}, the flux density { for the peak of the reverse shock emission at $\sim10$\,days post-merger} is $\sim10^{-7} - 10^{-5}$\,$\mu$Jy for a top-hat jet\footnote{The steep rise to peak in the top-hat case, and the peak duration for the off-axis observed afterglow is likely an effect of the afterglow approximation.
Where the full density profile of the forward shock is considered, the rise-time is earlier and the peak is broadened \citep[e.g.][]{vaneerten2012, decolle2012}. 2D hydrodynamic simulations have shown that the afterglow from a top-hat jet, when observed off-axis, will look more like a structured outflow \citep{gill2019}. 
In such a case, we expect the reverse shock to appear similar to the case of an angular structured outflow shown here} with an opening angle of $6^\circ$.
For the structured outflows a magnetization parameter $R_B>1$ is required {for the reverse shock to dominate the early afterglow}.
Where this is the case, the flux density is $\sim0.01 - 1$\,$\mu$Jy.
Observations at this level will be extremely difficult, however, { for rare events that have an intrinsically high energy and/or are very close, $\sim10$s of Mpc, these limits will be less restrictive.}

The change in the temporal index { may} reveal the reverse shock { where an afterglow that is detectable at early times}.
In Fig.~\ref{fig:alpha3GHz}, the 5\,GHz rise index evolution with time before the afterglow forward shock peak is shown for an observer at $\iota=[0,~18,~36]^\circ$ or $\sim[0,~3,~6]\times\theta_c$.
{ The effects of scintillation are shown with faint lines and where scintillation is present, the rise index information is lost.}
{ However, the temporal behaviour of $\alpha$, where scintillation is not considered, is interesting.
For an off-axis observer, }where $R_B=1$, the reverse shock can be seen for the top-hat and marginally for the power-law case at $t/t_{\rm p}\sim10^{-1}$.
Whereas for the two-component and Gaussian models, the rise index gradually flattens to the peak at $\alpha=0$ as expected from a forward shock.
For higher values of $R_B$, the reverse shock is more obvious -- briefly steepening the incline at $t/t_{\rm p}\lesssim10^{-1}$ before a shallower rise then becoming forward shock dominated and peaking at $t/t_{\rm p}=1${, where $t_{\rm p}$ is the time when the forward shock emission peaks; the index for the forward shock only case is shown as a thin line and is clear for the $\iota=0^\circ$ row where the reverse shock contributes}.

Variability in the rise index for the afterglow at higher inclinations { may} show the contribution from a reverse shock. 
{This variability will be complicated by any scintillation and observations may not be sensitive enough to detect changes due to the reverse shock, however, any modulation due to scintillation can be used to measure the size of the outflow \citep[e.g.][]{granot2014} putting constraints on the jet vs cocoon origin of an early afterglow detection.
{ The effects of scintillation on the 5\,GHz emission for a source at 100\,Mpc are shown in Fig. \ref{fig:alpha3GHz}. 
On-axis, scintillation does not obscure the changes due to the reverse shock but for higher inclination systems the effects of scintillation are much more problematic and even for rare events at $<50$\,Mpc the effects of scintillation will likely wash-out any useful information.
However, strong scintillation is only present at frequencies below a transition frequency, typically $\sim10$\,GHz and so observations above the transition frequency will be limited by weak scintillation only \citep[see][for a review]{granot2014}.
Fig. \ref{fig:alphaALMA} shows the rise index $\alpha$ for the same source as Fig. \ref{fig:alpha3GHz} but at an observed frequency of 97.5\,GHz.
The reverse shock emission is weaker at these higher frequencies (for our parameters), so we show only the cases for $R_B>1$.
The same characteristic changes in the rise index behaviour for the various jet structures can be seen.
From Fig. \ref{fig:THalma} it is clear that ALMA is not sensitive enough to detect the afterglow at the required time of the reverse shock for an inclined system, however, we note that at frequencies $>10$\,GHz the VLA has sensitivity comparable to that shown in Fig. \ref{fig:TH3GHz} and the effects of strong scintillation are similarly suppressed.
}

Interestingly, from Figs.~\ref{fig:alpha3GHz} and \ref{fig:alphaALMA}, where the temporal index for an on-axis observer appears similar for each of the jet structures during the rising afterglow, the post-peak decline reveals some differences where the structure of the outflow is extended.
In the off-axis cases, the difference post-peak is not obvious, however, pre-peak the behaviour of the index can be used to indicate the presence of structure.
For a top-hat jet, the index $\alpha$ is always much more variable than for any of the structured outflow cases.
Where the reverse shock is observed, the timescale on which it appears and the subsequent behaviour of the afterglow light-curve can be used to distinguish between a two-component structure and either a power-law or Gaussian structure -- this will require self-consistent modelling of both the reverse and the forward shock systems with broadband data.}

{Application of the reverse shock model to the afterglow of GW170817 is shown in Fig.~\ref{fig:GW170817}.
Here the radio data is $4-8$\,GHz from \cite{hallinan2017, margutti2018, mooley2018a, dobie2018} and the $R$-band data is from \cite{lyman2018, lamb2019}.
The afterglow model has parameters\footnote{Isotropic equivalent kinetic energy of the central jet core $E_{\rm K}=10^{52.4}$\,erg, central jet core Lorentz factor $\Gamma_0=427$, ambient density $n=10^{-3.52}$\,cm$^{-3}$, microphysical parameters $\varepsilon_B=10^{-3.11}$, $\varepsilon_e=10^{-1.14}$, $p=2.164$, and jet core angle $\theta_c=5.6^\circ$ and inclination $\iota=21.2^\circ$} consistent with the posterior distribution for a Gaussian structured jet from \cite{lamb2019} and the reverse shock where $R_B=[1,50,500]$ are shown as an excess at early times, $t\sim 1$\,day post-merger.
Scintillation at 5\,GHz is included for the sky localisation of GW170817 from NE2001 \citep{cordes2002} and is shown as a shaded region representing the maximum and minimum variability following \cite{granot2014}.}

For short GRBs at cosmological distances, the afterglow is expected to be at small inclinations within the jet opening angle or the core angle for a structured jet.
From Fig.~\ref{fig:TH3GHz} {and \ref{fig:THalma}} it is clear that on-axis the different structures show little difference -- {however, see Fig.~\ref{fig:alpha3GHz} where post-peak some difference could be apparent in the late afterglow decline phase}.
For untriggered transient surveys such as the Square Kilometre Array, the brighter-for-longer duration of a structured outflow with a reverse shock will increase the likelihood of making multiple detections of an orphan afterglow at higher distances.
For optical transient surveys such as the Large Synoptic Survey Telescope, the reverse shock is not expected to be bright and the transient rate estimated from forward shock considerations will remain unaffected by inclusion of a reverse shock \citep[e.g.][]{lamb2018c}.

\section{Conclusions}
\label{sec:conc}

For mildly inclined $10^\circ\lesssim\iota\lesssim30^\circ$ GW detected binary neutron star mergers within $\sim100$\,Mpc, the reverse shock will show a distinct feature in the rising afterglow emission at $0.1-10$\,days post-merger.
For structured outflows described by a power-law or Gaussian profile, the reverse shock will appear as an early bump or plateau before a gradual rise to peak at $\sim100$\,days.
{For a two-component structure the off-axis emission from the jet core will dominate for observers at $\sim2\times\theta_c$ but at inclinations $\iota\gtrsim3\times\theta_c$ then the reverse shock from the wider component will contribute where $R_B\sim500$ at a slightly later time than for the power-law or Gaussian structured case.
For a top-hat jet, where the jet has no angular structure, then the reverse shock will be fainter than the equivalent from a structured outflow and followed by a sharp rise to peak.} 
The flux density level of the reverse shock can be used to estimate the degree of magnetization within the outflow ejecta {where broadband observations of the afterglow will constrain the various parameters}.

\section*{Acknowledgements}

We thank the anonymous referee for helpful comments that have improved the paper.
GPL thanks Alexander van der Horst and Klaas Wiersema for useful discussions.
GPL is supported by STFC grant ST/S000453/1.




\bsp	
\label{lastpage}
\end{document}